\documentclass[12pt]{article}
\usepackage{epsfig}
\usepackage{amssymb}
\newcommand{\ba}{\begin{array}{c}}

\newcommand{\ea}{\end{array}}

\newcommand{\be}{\begin{equation}}
\newcommand{\ee}{\end{equation}}
\newcommand{\ben}{\[}
\newcommand{\een}{\]}

\newcommand{\kt}{\rangle}
\newcommand{\br}{\langle}

\newcommand{\bbr}{\br\!\br}

\newcommand{\pbr}{\prec\!}
\newcommand{\pkt}{\!\succ}

\begin{document}

\begin{center}

{\Large \bf

The coupled-cluster approach to quantum many-body problem in a
three-Hilbert-space reinterpretation

 }

\vspace{13mm}

 {\bf Raymond F. Bishop}

\vspace{3mm}

 {School of Physics and Astronomy, Schuster Building,
  The University of Manchester, Manchester, M13 9PL,
  UK\footnote{e-mail: raymond.bishop@manchester.ac.uk}}

\vspace{3mm}

 and

\vspace{3mm}

{\bf Miloslav Znojil}

\vspace{3mm}

 {
  Nuclear Physics Institute ASCR, 250 68 \v{R}e\v{z}, Czech
Republic\footnote{e-mail: znojil@ujf.cas.cz}}

\vspace{3mm}

\end{center}

\date{ }

 \noindent
The quantum many-body bound-state problem in its computationally
successful coupled cluster method (CCM) representation is
reconsidered. In conventional practice one factorizes the
ground-state wave functions $|\Psi\kt= e^S\,|\Phi\kt$ which live in
the ``physical'' Hilbert space ${\cal H}^{(P)}$ using an elementary
ansatz for $|\Phi\kt$ plus a formal expansion of $S$ in an operator
basis of multi-configurational creation operators $C^{+}_{\jmath}$.
In our paper a reinterpretation of the method is proposed. Using
parallels between the CCM and the so called quasi-Hermitian, {\it
alias} three-Hilbert-space (THS), quantum mechanics, the CCM
transition from the known microscopic Hamiltonian (denoted by usual
symbol ${{H}}$), which is self-adjoint in ${\cal H}^{(P)}$, to its
effective lower-case isospectral avatar
${{\hat{h}}}=e^{-S}{{H}}e^S$, is assigned a THS interpretation. In
the opposite direction, a THS-prescribed, non-CCM, innovative
reinstallation of Hermiticity is shown to be possible for the CCM
effective Hamiltonian ${{\hat{h}}}$, which only appears manifestly
non-Hermitian in its own (``friendly'') Hilbert space ${\cal
H}^{(F)}$. This goal is achieved via an {\em ad hoc} amendment of
the inner product in ${\cal H}^{(F)}$, thereby yielding the third
(``standard'') Hilbert space ${\cal H}^{(S)}$. Due to the resulting
exact unitary equivalence between the first and third spaces, ${\cal
H}^{(P)}\sim {\cal H}^{(S)}$, the indistinguishability of
predictions calculated in these alternative physical frameworks is
guaranteed.

\vspace{5mm}

\vspace{2mm}




\newpage

\section{Introduction}


The coupled cluster method (CCM) of construction, say, of the
ground-state energies and wave functions of general quantum
many-body systems works with virtual multi-particle excitations, and
the linked-cluster nature of the contributions to the resulting
estimates of measurable quantities is particularly emphasized
\cite{bishop1} -- \cite{bishop3}. The strategy leads, in practical
calculations, to the replacement of a given, known, realistic and
exact microscopic input Hamiltonian (let us denote it by the
dedicated symbol ${{H}}$) by its lower-case isospectral
reparametrization
 \be
 {{\hat{h}}}=\Omega^{-1}\,{{H}} \Omega\,.
 \label{reparama}
 \ee
An optimal similarity-mediating transformation operator $\Omega$ is
then sought in an exponential, manifestly linked-cluster form
$\Omega= \exp S$. The excitations themselves are usually assumed
multi-configurational, multi-indexed and generated by a complete set
of mutually commuting many-body creation operators $C^{+}_{{\jmath}}
\equiv (C^{-}_{{\jmath}})^{\dagger}$ such that, conventionally,
$C^{+}_{0} \equiv I$ and $ C^{-}_{0}\equiv I$ while  $ S =
\sum_{{\jmath}\neq 0}{\cal S}_{{\jmath}}C^{+}_{{\jmath}}$.

Naturally, the quality of the variationally determined CCM
coefficients ${\cal S}_{{\jmath}}$ translates into the quality of
the predicted expectation values of any operator of an observable
quantity. In practice, there expectedly emerges a conflict between
the precision and the costs of the results. One is thusforced to
find an optimal compromise between these two requirements by
introducing various approximation schemes. In our present short
paper we intend to describe one  possible systematic approach to the
abstract formulation of approximation hierarchies.

Our considerations will be inspired by the recent progress achieved
in both the formal and the  applied analyses of isospectral
partnerships ${{\hat{h}}} \leftrightarrow {{H}}$. In particular, we
shall emphasize the innovative role  played by various {\em
non-unitary} mappings $\Omega$, say, in their alternative
time-independent or time-dependent forms as described in review
papers \cite{Geyer} and \cite{SIGMA}, respectively.

Once a decisive simplification of the Hamiltonian is achieved by a
{non-unitary} map $\Omega: {{H}} \to {{\hat{h}}}$, we have to start
working with the less usual form ${{\hat{h}}}$ of the Hamiltonian
which becomes, in general, {non-Hermitian} since
 \be
 {{\hat{h}}}^\dagger =
 \Omega^\dagger {{H}} \left (\Omega^{-1}\right )^\dagger=
 \Omega^\dagger \Omega\,{{\hat{h}}}\,\Omega^{-1} \left (\Omega^{-1}
 \right )^\dagger\,, \ \ \ \ \ \
 \Omega^\dagger\Omega\ \equiv \ \Theta \neq I
 \,.
 \label{prehe}
  \ee
In our present paper we intend to reveal and describe a deeper
relationship between the CCM and the abstract framework provided by
the mathematical theory of Hamiltonians exhibiting the above
property of quasi-Hermiticity \cite{Dieudonne}, {\em alias}
crypto-Hermiticity \cite{Smilga}, with respect to the alternative
Hilbert-space metric-operator $\Theta\neq I$,
 \be
 {{\hat{h}}}^\dagger \Theta =\Theta\,{{\hat{h}}}
 \,.
 \label{dieudonne}
 \ee
In section \ref{ths-formalism} we shall explain the abstract
formalism of three-Hilbert-space (THS) representation of quantum
systems. We shall make use of the notation conventions of review
paper~\cite{SIGMA}, however, with the single, CCM-adapted exception
of an interchange of the meaning of the lower- and upper-case
symbols for the Hamiltonian. For the sake of clarity,
Table~\ref{tab:widea} offers the explicit translation of the present
notation conventions (as displayed in the first column) to the
language of Ref.~\cite{SIGMA} (given in the second column).
Subsequently, in section \ref{thsaccm-formalism} an overall review
of the key ideas of CCM constructions will be recalled, and their
reinterpretation within the general THS scheme will be described.
Section \ref{summary} will finally summarize our observations and
proposals.

\section{THS representation of a quantum system\label{ths-formalism}}

\begin{table*}
\centering
\begin{tabular}{||c||c||c||}
 \hline
 \hline
  {\bf \ \ \ concept }
  &{\bf \ \ \ \ \ \ CCM, ref.  \cite{bishop3}}
  &
  {\bf \ \ \  \ \ \ THS, ref.  \cite{SIGMA}}  \\
  \hline
  \hline
  (realistic, microscopic)
  &  (Hermitian)
  &   (Hermitian in ${\cal H}^{(P)}$)\\
  initial Hamiltonian
  &
  $H$ & $\mathfrak{h}$ \\
  \hline
  (non-unitary)
  &
   (creation) & (general invertible map) \\
  transformation
  &
  $\exp S$ & $\Omega:{\cal H}^{(F)}\to {\cal H}^{(P)}$\\
   \hline
   (assumed simplified) & (non-Hermitian)
  &  $ \left \{
  \ba {\rm non\!\!-\!\!Hermitian\  in \ } {\cal H}^{(F)}
  \\
    {\rm  and \ Hermitian \ in\ }{\cal H}^{(S)}
     \ea
     \right \}
     $\\
 transformed Hamiltonian
  &
  $\hat{h}=e^{-S}He^S$ & $\mathfrak{H}=\Omega^{-1}\mathfrak{h}\Omega$\\
          \hline
   \hline
\end{tabular}
\caption{Warning: opposite notation conventions.} \label{tab:widea}
\end{table*}

\subsection{Inspiration: Fourier transform}

The most elementary one-dimensional harmonic-oscillator Hamiltonian
 \ben
 {{H}}^{(HO)} = -\frac{d^2}{dx^2} + x^2
 \label{HO}
 \een
may be recalled as one of the best known examples of an operator
representing a typical quantum observable. It enters the ordinary
differential Schr\"{o}dinger equation
 \be
 {{H}}^{(HO)}\,\psi_n^{(P)}(x) = E_n^{(HO)}\,\psi_n^{(P)}(x)
 \,, \ \ \ \psi_n^{(P)}(x) \in  L^2(\mathbb{R})
 \,, \ \ \ n=0,1,\ldots\,
 \label{SEho}
 \ee
for ``physical'' wave functions $\psi_n^{(P)}(x)$. The solution of
this eigenvalue problem yields the well known discrete spectrum of
bound-state energies $E_0=1$, $E_1=3$, $E_2=5$, \ldots\,, while the
related wave functions belong to the most common Hilbert space of
square-integrable complex functions of $x \in \mathbb{R}$. The
argument $x$ of the wave functions coincides with an admissible
value of the position of the quantum particle in question. In other
words, the $^{(P)}-$superscripted complex functions
$\psi_n^{(P)}(x)$ may be interpreted as yielding the probability
density of finding the particle at point $x \in \mathbb{R}$.

The wave functions in question live in a physical Hilbert space $
L^2(\mathbb{R})\,\equiv \,{\cal H}^{(P)}$. Formally, these functions
may be represented as Fourier transforms of elements of a,
supposedly, ``friendlier'' Hilbert space, $\psi_n^{(P)}={\cal F}
\psi_n^{(F)}$,  $\psi_n^{(F)} \in {\cal H}^{(F)}$. By construction,
the latter space is also $L^2(\mathbb{R})$ but the physical meaning
of the argument $p \in \mathbb{R}$ of the new wave functions
$\psi_n^{(F)}(p)$ is different. At the same time, the primary
observable (i.e., the energy) remains unchanged.

In practice, the harmonic oscillator appears equally well
represented in {\em both} of the Hilbert spaces ${\cal H}^{(P,F)}$.
%
%
%
Whenever one moves to a more complicated model, however, one may
find that one of these spaces is preferable.
%
%
%
In other words, a unitary-mapping-mediated transition to a
potentially friendlier Hilbert space ${\cal H}^{(F)}$ should be
employed whenever it appears to lead, say, to a simplification of
the calculation of the energies or of the wave functions.

We only have to add here that the same recommendation remains valid
even for mappings ${\cal H}^{(P)}\leftrightarrow {\cal H}^{(F)}$
which cease to be unitary. In this sense, our freedom of choosing
between the upper- and lower-case Hamiltonians  as expressed in
Eq.~(\ref{reparama}) may prove important, say, as a source of
acceleration of the rate of convergence of various numerical or
variational calculations \cite{Geyer}.

%

\subsection{Non-unitary mappings $\Omega=\exp S$}

Our present text is basically inspired by the recent  growth of
popularity of quantum models in which the {\em ad hoc} non-unitary
isospectral transformations
 \be
 {{H}} \to {{\hat{h}}}=\Omega^{-1}\,{{H}}\,\Omega\,
 \label{tady}
 \ee
perceivably simplify the Hamiltonian. Thus, Eq.~(\ref{tady}) offers
a path towards the feasibility of the evaluation of  bound-state
energies in complicated quantum systems via an $\Omega-$mediated
transition from a complicated ``primary'' Hilbert space ${\cal
H}^{(P)}$ to a ``friendlier'' Hilbert space ${\cal H}^{(F)}$.

\subsubsection{Crypto-Hermitian IBM method.}

One should distinguish between several non-equivalent applications
of the above-outlined ideas. In one of the key references on the
whole subject \cite{Geyer} the authors start from the knowledge of
an overcomplicated ${{H}}$ and from a qualified guess of a suitable
simplification mapping $\Omega\neq \left ( \Omega^\dagger\right
)^{-1}$. For a persuasive illustration of the practical efficiency
of such an approach the authors recalled the so-called
interacting-boson-model (IBM) calculations of the spectra of heavy
atomic nuclei. Using the Dyson-Maleev choice of the boson-fermion
mappings $\Omega^{\rm (Dyson)}$ this strategy was found to lead to
successful and particularly computation-friendly forms of
variational predictions of the measured energy levels \cite{Jolos}.

The key condition of applicability of the latter IBM recipe may be
seen in the feasibility of construction of the ultimate
``effective'' Hamiltonian ${{\hat{h}}}$ of Eq.~(\ref{tady}). One
arrives at a non-Hermitian operator in general, ${\hat{h}}\neq
{\hat{h}}^\dagger$. It is worth adding that an exception may occur
when the original self-adjoint Hamiltonian ${{H}}$ accidentally
happens to commute with the operator-product symmetry $\Pi=\Omega\,
\Omega^\dagger$; notice that $\Pi\neq \Theta$ unless we restrict
attention to the mere normal-operator mappings $\Omega$ such that
$\Omega^\dagger\Omega=\Omega\Omega^\dagger$.

Whenever ${\hat{h}}\neq {\hat{h}}^\dagger$, the practical
determination of the eigenvalues of the transformed Hamiltonian must
remain easy and efficient. The reason is that in comparison with
standard methods, one must replace the usual single time-independent
Schr\"{o}dinger equation by the following {\em doublet} of conjugate
eigenvalue problems
 \begin{equation}
 {{\hat{h}}}|\Phi_n\rangle = E_n|\Phi_n\rangle\,,\ \ \ \
  \langle\tilde{\Phi}_m|{{\hat{h}}} =
 E_m\langle\tilde{\Phi}_m|\,, \ \ \ n,m = 0, 1, \ldots\,
 \label{eq:SE}
 \end{equation}
using the respective action-to-the-right and action-to-the-left
conventions.

Interested readers may  consult review paper \cite{SIGMA} in which a
detailed discussion of further subtleties is given, first of all,
for the far from trivial Heisenberg-representation-like cases in
which the non-unitary mapping $\Omega$ is also permitted to vary
with time.

\subsubsection{${\cal PT}-$symmetric models.}

A reversal of the application of the simplification ${{H}}\to
{{\hat{h}}}$ may be found promoted in the overall context of
relativistic quantum field theory. In this entirely different domain
of physics, Bender and his coauthors were the first who advocated an
alternative philosophy of first choosing a sufficiently elementary
non-Hermitian ${{\hat{h}}}$ and of postponing the reconstruction of
the overcomplicated selfadjoint operator ${{H}}$, sometimes even
indefinitely.

The initial move is due to to Bender and Boettcher who published, in
1998, an influential letter \cite{BB}. In this work they noticed
that certain elementary non-Hermitian toy-model operators
${{\hat{h}}}$ appeared to possess real and bound-state-like spectra,
which were discrete, non-degenerate and bounded from below. In 2001,
their observations were rigorously proved while, a few years later,
some of these results were also complemented by approximate
reconstructions of the necessary metric operator(s)
$\Theta=\Theta({{\hat{h}}})$ (cf., e.g., review \cite{Carl} for
details).

On a model-independent level these developments finally resulted in
a fully consistent innovative THS strategy in which one {\em starts}
from a sufficiently elementary lower-case (i.e., non-Hermitian)
candidate for a ``realistic-model'' Hamiltonian ${{\hat{h}}}\neq
{{\hat{h}}}^\dagger$. Under a number of assumptions (cf., e.g.,
reviews \cite{Geyer,ali1,ali,ali3}) one is then able to {\em
re-construct} a suitable Hilbert-space mapping
$\Omega=\Omega({{\hat{h}}})$ and, via Eq.~(\ref{reparama}), also a
self-adjoint, textbook-compatible isospectral avatar
${{H}}={{H}}^\dagger$ of the Hamiltonian living in ${\cal H}^{(P)}$.
In other words, from the initial knowledge of a
quantum-dynamics-determining operator ${{\hat{h}}}$ one is able to
reconstruct, in principle at least, one or several tractable,
textbook-compatible phenomenological quantum-mechanical and/or
field-theoretical models.

Naturally, the initial choice of  Hamiltonian ${{\hat{h}}} \neq
{{\hat{h}}}^\dagger$ acting in ${\cal H}^{(F)}$ should guarantee
that the pair of Schr\"{o}dinger Eqs.~(\ref{eq:SE}) remains
sufficiently easily solvable. This requirement is not so easily
satisfied. In practice people usually accept various independent and
additional simplification assumptions, therefore. Among them, a
truly exceptional status belongs to the so called ${\cal
PT}-$symmetry assumption or, more generally, to the assumption of
the so called pseudo-Hermiticity property of ${{\hat{h}}}$
(interested readers should consult, e.g., review \cite{ali} for more
details).

\subsubsection{Towards the complex energy spectra.}

Another, third and still different implementation of the
non-Hermitian-observable ideas is much older than the previous two.
It may be traced back to the traditional model-space projection
technique of Feshbach in which one of the non-unitary mappings
$\Omega$ and $\Omega^{-1}$ is chosen as a projector so that the
other one cannot exist. It is well known that the resulting
simplified effective Hamiltonians are restricted to a subspace while
becoming energy-dependent in general. In this sense, Feshbach's
effective Schr\"{o}dinger Eqs.~(\ref{eq:SE}) are {\em de facto}
nonlinear.

Such a case certainly lies outside the scope of our present
considerations.  Still, it is worth noting that recently, there has
emerged a number of papers in which the authors pointed out the
existence of numerous links between the latter studies of resonances
(i.e., of the quantum Hamiltonians possessing complex spectra) and
their above-mentioned real-spectrum alternatives. Interested readers
may consult, e.g., monograph \cite{Nimrod} to see a number of newly
discovered connections between the physics of Hermitian and/or
non-Hermitian effective Hamiltonians and the related mathematics
which recommends, say, the use of the concepts of the Kato's
exceptional points, etc.

One should also point out that even in the recent physics-based and
experiment-oriented studies of the real-spectrum pseudo-Hermitian
and ${\cal PT}-$symmetric models models there has been a definite
increase of interest in the interdisciplinary applications of the
THS-related concepts of the spontaneous ${\cal PT}-$symmetry
breakdown and/or explanations of the exceptional-point-related
phase-transition mechanisms connected with the loss of the reality
of the spectrum (cf., e.g., the recent quantum-theory-related review
paper \cite{Ingrid}, or a sample \cite{Makris} of a successful
transfer of these ideas even beyond the realm of quantum theory
itself).

\section{\label{thsaccm-formalism}THS interpretation of CCM
constructions}


Having passed through the extensive list of motivating
considerations we are now getting very close to the key purpose of
our present paper. For the construction of a concrete and
``optimal'' mapping $\Omega=\Omega({{H}})$ in the THS context we see
that we might accept directly some of the CCM constructive
techniques. {\em Vice versa}, in the CCM framework we encounter the
possibility of extending its philosophy and its range beyond the
ground-state constructions. For this purpose we may decide to
experiment with various THS-inspired alternatives to the basic
(bi-)variational CCM ans\"{a}tze.

In an introductory step let us return, therefore, to the
IBM-motivated version of the THS approach in which one assumes a
full knowledge of the realistic, albeit prohibitively complicated,
Hamiltonian ${{H}}=H^\dagger$, defined in some microscopic physical
Hilbert space ${\cal H}^{(P)}$. A qualified guess or construction of
$\Omega$ will be then vital for the success of computations, i.e.,
first of all, for the success of the practically tractable
construction and solution of the pair of Schr\"{o}dinger
Eqs.~(\ref{eq:SE}).


\subsection{\label{ccm-formalism}Brief introduction to
CCM constructions}

In the CCM context, the generic, Dyson-inspired non-unitary mapping
$\Omega^{(CCM)}$ has traditionally been considered in the specific
linked-cluster form of an exponential operator $\Omega^{(CCM)}=\exp
S$. In the literature  (cf., e.g., Ref.~\cite{Bi:1998_LectPhys_V510}
with further references) one may find a huge number of practical
applications of the CCM strategy by which the ground-state wave
functions are sought in the form of products
 \be
 |\Psi\kt= e^S\,|\Phi\kt\,.
 \label{CCMans}
 \ee
The ket vector $|\Phi\rangle$ represents here a normalized state
(usually called the model state or reference state), intended to be
employed as a cyclic vector with respect to a complete set of
mutually commuting multi-configurational creation operators
$C^{+}_{\jmath} \equiv (C^{-}_{\jmath})^{\dagger}$. Our use of the
special symbol $\jmath$ for the index indicates that this is a
multi-index that labels the set of all many-particle configurations.
In other words, states of the many-particle quantum system in
question can be all written as superpositions of basis states
$C^{+}_{\jmath}|\Phi\rangle$.

Variational eigenkets (\ref{CCMans}) of the many-body self-adjoint
Hamiltonian ${{H}}={{H}}^\dagger$ are conveniently written in terms
of the specific CCM {\em operator} ansatz
\begin{equation}
 S =
  \sum_{\jmath\neq 0}{\cal S}_{\jmath}C^{+}_{\jmath}\,.   \label{eq:ket_eq}
 \end{equation}
The fundamental CCM replacement (\ref{CCMans}) of an unknown vector
$|\Psi\kt$ by an unknown operator $S$ is very well motivated from
several independent points of view. One of the motivations is
inherited from Rayleigh-Schr\"{o}dinger perturbation theory in
which, at certain stage of construction, the {\em operator}
Schr\"{o}dinger equation $H|\Psi\kt=E|\Psi\kt$ in question is
replaced by its {\em single} bra-vector projection $\br 0|
H|\Psi\kt=E\br 0|\Psi\kt$ or, more generally, by a {\em finite}
multiplet of such projections $\br 0_j| H|\Psi\kt=E\br 0_j|\Psi\kt$.

The key advantage of such a reduction lies in the possibility of a
variationally optimal choice of the bra-vectors $\br 0_j|$. By
contrast, the property of the Hermiticity of Hamiltonian ${{H}}$
becomes, to a large degree, irrelevant. Thus, one transfers this
experience to the CCM context by introducing a complementary,
formally redundant concept of left-action variational eigenvector
$\langle\tilde{\Psi}|$ of $H$.

The nontrivial difference between the tilded and untilded
eigenvector  $|\tilde{\Psi}\rangle$ and $|\Psi\rangle$ is motivated
by the possibility of introduction of an additional set
$\{\tilde{{\cal S}_{\jmath}}\}$ of free parameters in the bra-vector
\begin{equation}
\langle\tilde{\Psi}| = \langle\Phi|\tilde{S}\mbox{e}^{-S}; \qquad
\tilde{S} = I + \sum_{\jmath\neq0}\tilde{{\cal
S}_{\jmath}}C^{-}_{\jmath}. \label{eq:bra_eq}
\end{equation}
Together with the conditions of completeness of the basis
\begin{equation}
\sum_{\jmath}C^{+}_{\jmath}|\Phi\rangle \langle\Phi|C^{-}_{\jmath} =
I = |\Phi\rangle \langle\Phi| + \sum_{\jmath \neq
0}C^{+}_{\jmath}|\Phi\rangle \langle\Phi|C^{-}_{\jmath}\,,
\end{equation}
and together with the usual properties of the creation and
annihilation operators,
\begin{equation}
C^{-}_{\jmath}|\Phi\rangle = 0 = \langle\Phi| C^{+}_{\jmath}; \qquad
\forall {\jmath} \neq 0
\end{equation}
and
\begin{equation}
[C^{+}_{\jmath},C^{+}_{J}]=0=[C^{-}_{\jmath},C^{-}_{J}]
\label{commutation}
\end{equation}
we arrive at the standard version of the CCM formalism in which one
currently employs approximations which do not make use of the
manifest Hermiticity of the original eigenvalue problem. Such
approximations may entail keeping only a physically motivated subset
of the multi-indices $\jmath$ in the otherwise exact expansions of
the correlation operators $S$ and $\tilde{S}$ in Eqs.~(\ref{CCMans})
- (\ref{eq:bra_eq}).

As an immediate mathematical consequence, the CCM Schr\"{o}dinger
equation for ground state acquires the two different and mutually
non-conjugate alternative forms
 \begin{equation}
 {{\hat{h}}}\,|\Phi\rangle = E |\Phi\rangle\,,
 \ \ \ \ \ \
 \langle\Phi|\tilde{S}\,{{\hat{h}}}=E\langle\Phi|\tilde{S}\,,
 \ \ \ \ \ \
 {{\hat{h}}}=\mbox{e}^{-S} {{H}} \mbox{e}^{S}\,.
 \label{eq:CCM_SE}
 \end{equation}
Obviously, once the two sets of coefficients $\{{\cal S}_{\jmath}\}$
and $ \{\tilde{\cal S}_{\jmath}\}$ are determined, all the
ground-state properties of the many-body system in question may be
considered as known.

The ground-state expectation value of any given operator $\Lambda$
should be evaluated from the asymmetric prescription
\begin{equation}
\langle\tilde{\Psi}|\Lambda\,|\Psi\rangle =
\langle\Phi|\tilde{S}\mbox{e}^{-S}\Lambda\mbox{e}^{S}|\Phi\rangle =
\bar{\Lambda}({\cal S}_{\jmath}, \tilde{{\cal S}_{\jmath}})\,.
\label{bar_A}
\end{equation}
This recipe keeps trace of the artificial asymmetry as introduced in
Eq.~(\ref{eq:CCM_SE}) which, in its turn, simplifies certain
technical aspects of the global CCM approach. In particular, in the
bi-variational spirit the energy expectation formula
 \be
 \langle\tilde{\Psi}|{{H}}|\Psi\rangle=\langle\tilde{\Phi}|
\hat{h} |\Phi\rangle
 \ee
may now be minimized with respect to the full set of parameters
$\{{\cal S}_{\jmath},{\tilde{\cal S}}_{\jmath}\}$. Two equations
follow, viz.,
\begin{equation}
\langle \Phi|C^{-}_{\jmath}{{\hat{h}}}|\Phi\rangle = 0\,; \ \ \ {\rm
} \ \ \ \
  \forall \jmath \neq 0    \label{eq:ket-coeff}
\end{equation}
and
\begin{equation}
 \langle\Phi|\tilde{S}({{\hat{h}}} - E)C^{+}_{\jmath}|
 \Phi\rangle = 0\,; \ \ \  {\rm } \ \ \ \ \forall \jmath \neq 0.
 \label{eq:Bra-coeff-1}
\end{equation}
In their turn, these relations may be interpreted as a coupled
algebraic set of equations that determine the parameters $\{{\cal
S}_{\jmath},{\tilde{\cal S}}_{\jmath}\}$. The consistency of the
recipe may be reconfirmed by the derivation of the former relation
(\ref{eq:ket-coeff}) from the assumption of completeness of the set
of states $\{\langle\Phi|C^{-}_{\jmath}\}$. Similarly,
Eq.~(\ref{eq:Bra-coeff-1}) may be perceived as a consequence of the
completeness of the conjugate set $\{C^{+}_{\jmath}|\Phi\rangle\}$.


The coupled equations (\ref{eq:ket-coeff}) and
(\ref{eq:Bra-coeff-1}) are of the Goldstone linked-cluster type. For
this reason, all extensive variables, such as the energy,  scale
linearly with the  number of particles at every level of
approximation. This is another merit of the CCM construction. Among
the disadvantages we mention that the ground-state energy formula
does not necessarily provide an upper bound, due to the intentional
violation of manifest Hermiticity for the problem. Still, the recipe
enables us to determine both the quickly convergent energies as well
as the Hamiltonian-dependent values of parameters $\{{\cal
S}_{\jmath},{\tilde{\cal S}}_{\jmath}\}$ or, in various approximate
schemes, of the respective truncated subsets of these values.

%

Within the general framework of the CCM treatment of many-body
quantum systems some of the above-mentioned assumptions and
restrictions may be removed. The method may certainly be extended,
say, to cover also excited states and/or certain time-dependent
versions of dynamics. In both of these directions, an implementation
of ideas from THS context might prove particularly helpful.

%

\subsection{CCM - THS correspondence}

The close mathematical relationship between the various variational
CCM recipes and the universal three-Hilbert-space (THS)
representation of a generic quantum system has been largely
overlooked till now. Apart from a few rather inessential
differences, one of the key obstacles may be seen in the differences
in their notations, a first sample of which is displayed in Table
\ref{tab:widea} where we see that for Hamiltonians, the CCM and THS
notation conventions are strictly opposite (so we have to
re-emphasize that in our present paper we are using the first-column
notation conventions).

With the due care paid to the Hermiticity or non-Hermiticity of the
Hamiltonian, it seems equally important to spot the CCM - THS
coincidences and/or differences in the definitions and meanings of
the other concepts. For the ground-state wave functions, in
particular, the parallels in the denotation of the same feature or
quantity are displayed in Table \ref{tab:widebe}.

\begin{table*}
\centering
\begin{tabular}{||l||l|l||}
 \hline
 \hline
  {\bf \ \ \ ground state }
  &{\bf \ \ \ \ \ \  CCM, ref.   \cite{bishop3}}
  &
  {\bf \ \ \ \ \ \ THS, ref.
 \cite{SIGMA}}  \\
    \hline
  \hline
   purpose: &
  bi-variationality
    & re-Hermitization of $H$ in ${{\cal H}^{(S)}}$\\
  \hline
   assumptions: &
  $\tilde{S}$ = annihilation
    &   $\Theta=\Omega^\dagger\Omega$,  $\ \ \ \ \ \Omega$ = invertible\\
    \hline
  \hline
    eigen-ket (simplified)
  &
  $|\Phi\kt$& $|0\kt \in {\cal H}^{(F,S)}$\\
 eigen-bra (conjugate)
  &
  $\br \Phi|$& $\br 0| \in {{\cal H}^{(F)}}'$\\
 eigen-bra (amended)
  &
  $\br \tilde{\Phi}|:=\br \Phi|\tilde{S}$
    & $\bbr 0| := \br 0|\Theta \ \in {{\cal H}^{(S)}}'$\\
   microscopic  ground state
  &
  $|\Psi\kt$& $|0\pkt \  =\Omega|0\kt \in {\cal H}^{(P)}$\\
   first variational ansatz &
  $\ \ \ \ \ \ \  =e^S|\Phi\kt$&  \\
  left ground state
  &
  $\br \tilde{\Psi}|$& $\pbr 0|\  =\br 0|\Omega^\dagger = \bbr
  0|\Omega^{-1}
  \in {{\cal H}^{(P)}}'$\\
  second variational ansatz
  &
  $\ \ \ \ \ \ \  =\br \tilde{\Phi}|e^{-S}$&
  \\
     \hline
   \hline
\end{tabular}
\caption{Parallel notation conventions.} \label{tab:widebe}
\end{table*}


An inspection of Table \ref{tab:widebe} reveals that in their
respective current versions, the two formalisms are far from
equivalent, indeed. At the same time, they may be both found to
suffer of certain specific weak points. In fact, our present
considerations were originally motivated precisely by a {\em
parallel} analysis of these respective weaknesses. After their
deeper study we came to the conclusion (documented and emphasized
also by the above compact review) that a perceivable profit might be
gained by modifying and getting those two formalisms and/or methods
of calculation closer to each other.

On the side of the CCM formalism, for example, one may immediately
notice an obvious contrast between the exponential CCM form of the
mapping $\Omega^{(CCM)}=\exp S^{(CCM)}$ and the manifestly {\em
non-exponential}, polynomial form of the tilded operator $\tilde{S}$
entering the second CCM ansatz (\ref{eq:bra_eq}). Naturally, such a
striking difference did not stay unnoticed in the related
literature, and the idea has been implemented into the so called
extended version of the CCM (ECCM) formalism \cite{bishop1,bishop2}.

On the side of the general THS formalism, in parallel, we may now
recollect one of the very popular formalism-simplifying tricks by
which one works just with the special Hermitian mappings
$\Omega_s=\Omega^\dagger_s=\exp S_s$ \cite{ali1,ali}. Under this
additional assumption one arrives at a fairly natural exponential
form of the equally special but still sufficiently general subset of
the positive-definite metrics, $\Theta_s=\exp 2S_s$. In this manner,
after the respective replacements $\tilde{S} \to \tilde{S}^{(ECCM)}$
and $\Theta \to \Theta_s=\exp 2S_s$, the initially very different
forms of the operators get closer.

Once one stops feeling discouraged by the similar, more or less
purely formal differences, one has to reopen also the question of
the respective roles of the operators $\tilde{S}$ and $\Theta$ in
the purely numerical context. This is another type of difference
which is, naturally, strongly dependent on the purpose of the
calculation. Traditionally, the CCM and THS calculation purposes are
truly rather different. Nevertheless, on the CCM side one
immediately notices that the predominance of the calculations of the
ground-state characterstics does not exclude extensions, say, to the
excited-state problem \cite{bishop4} or even to the description of
systems which are allowed to exhibit a manifest time-dependence of
their dynamics \cite{bishop5}. In this sense we are getting still
closer to the respective time-independent and time-dependent
non-Hermitian versions of the general and universal THS formulation
of abstract quantum mechanics as summarized, say, in
Refs.~\cite{Geyer} and \cite{SIGMA}.

\section{Discussion\label{summary}}

\subsection{A CCM - THS fusion?}

In the language of mathematics the core of our present message may
be summarized as follows: in fact, it need not be particularly
difficult to search for a further enhancement of parallels between
the manifestly non-Hermitian, annihilation-operator-type CCM choice
of the tilded operator $\tilde{S}$ and the strictly Hermitian and,
in addition, also strictly positive definite Hilbert-space-metric
operator $\Theta=\Omega^\dagger\Omega$. In the  terminology of
physics this persuasion is supported by the observation that what is
{\em shared} by both the {\em abstract} CCM and THS formalisms is a
truly exciting idea of using {\em nontrivial} ``redundant''
operators $\tilde{S}$ or $\Theta$ in place of the common {identity}
operator.

In both formalisms, the rationale behind the use of the respective
nontrivial operators $\tilde{S}$ and $\Theta$ is rather subtle
though fairly persuasive and not too dissimilar. Indeed, one starts
from a well known while, unfortunately, prohibitively complicated
initial self-adjoint Hamiltonian in both cases (recall, once more,
Table \ref{tab:widea}). Secondly, the choice and/or construction of
the mapping $\Omega=\exp S$ is motivated, in both of the approaches,
by a more or less comparably successful {\em simplification} of the
Schr\"{o}dinger eigenvalue problem. Thirdly, both the CCM and THS
re-arrangements of the quantum bound-state problem  lead to the
necessity of the introduction  of the respective nontrivial
operators $\tilde{S}$ and $\Theta$ using {\em comparably strong}
but, at the same time, {\em different} supportive arguments.

What now remains opens a truly challenging question is whether, and
in which sense, one could really achieve a {\em complete}
coincidence of the respective (and, apparently, ideologically
distant) CCM and THS recipes. Firstly, an affirmative answer may be
given (and the idea may be made working) whenever the Hilbert spaces
of the system remain, for whatever reason (i.e., e.g., for
approximation purposes) finite-dimensional.

In such a very specific case the space for a compromise immediately
opens after we move from the abstract formalism to any kind of a
practical variational calculation and/or numerical approximation.
Schematically speaking, any $2M-$parametric array of the
multi-indexed CCM variational coefficients ${\cal S}_{{\jmath}_k}$
and ${\tilde{\cal S}}_{{\jmath}_k}$ with $k=1,2,\ldots,M$ may be
perceived equivalent to an introduction of a $2M-$parametric metric
$\Theta=\Omega^\dagger\Omega$. It should be noted, as a supportive
argument, that even in the thorough IBM review \cite{Geyer} a large
amount of space has been devoted to the study of finite-dimensional
models and to the questions of practical variational applicability
of the THS scheme.

On this level of mathematics the overall nature and structure of the
above-indicated possibility of a complete unification (or, at least,
of a strengthening of the CCM-THS parallelism) may be read out of
the following three-Hilbert-space diagram,
 \ben
 \\
  \ba
    \ \ \ \ \ \ \ \ \begin{array}{|l|}
 \hline
   initial,\ given\ {\rm  microscopic\ Hamiltonian\ }
   {{H}}= {{H}}^\dagger\
   lives\ in \\
   {\rm  {\bf primary\ }  space\ }
    {\cal H}^{(P)};\ all\ is\ {\bf prohibitively\ complicated\!:\!}\\
 \ \ \ \bullet\ {\it one\  constructs}\  {\rm the
  \ CCM\ operator\ } \Omega=\exp\,S\\
 \hline
 \ea
  \ \ \ \ \ \ \ \ \ \ \  \  \  \ \ \ \ \ \
  \ \ \ \ \ \ \ \ \ \ \ \ \ \ \ \ \ \ \  \  \  \ \ \ \ \ \
  \ \ \ \ \ \ \ \
 \\
 \\
  %
 \stackrel{{\bf CCM\  map} \  \Omega^{-1}\, = \ \exp\,(-S)}{}
 \ \
  \swarrow\ \  \ \ \ \ \ \ \ \ \ \ \ \ \ \ \ \ \ \ \ \ \ \ \ \
 \ \ \ \ \  \searrow \nwarrow\
 \stackrel{\bf   equivalence}{} \ \ \ \ \ \ \ \
  \ \ \  \  \  \ \ \ \ \ \ \ \ \ \ \ \ \ \
  \ \ \ \ \ \ \ \ \ \ \ \ \ \ \ \ \ \ \  \  \  \ \ \ \ \ \
  \ \ \ \ \ \ \ \ \\
  \\
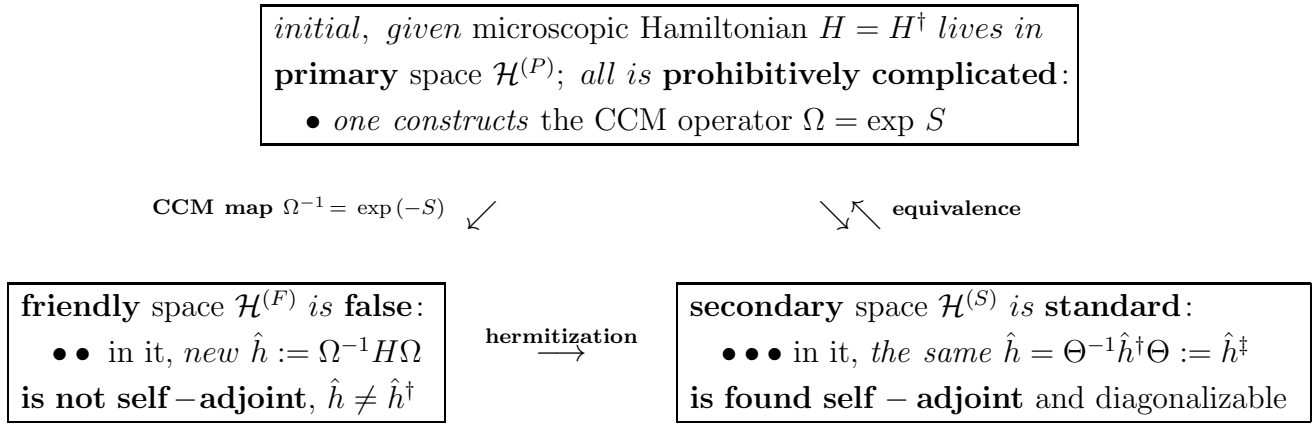
 \begin{array}{|l|}
 \hline
    {\rm   {\bf friendly \ } space} \
    {\cal H}^{(F)}\ is\ {\bf false\!:\!} \\
   \ \ \ \bullet  \bullet\ {\rm in\ it,}\ new\  {{\hat{h}}}
   :=\Omega^{-1}{{H}}\Omega \  \\
   {\rm  \bf is \ not\  self\!-\!adjoint,\ }{{\hat{h}}}
    \neq {{\hat{h}}}^\dagger\\
  \hline
 \ea
 \stackrel{ \ \ \ {\bf  hermitization} \ \ \  }{ \longrightarrow }
 \begin{array}{|l|}
 \hline
   {\rm  \bf secondary} \ {\rm  space\ } {\cal H}^{(S)}\
     is\ {\bf standard\!:\!}\\
   \ \ \ \bullet  \bullet  \bullet\
   {\rm in\ it, } \ {\it the\ same}\  {{\hat{h}}}
   =\Theta^{-1}{{\hat{h}}}^\dagger\Theta
   :={{\hat{h}}}^\ddagger {\rm \ }\\
 {\rm {\bf is\  found\  self-adjoint} \ and \ diagonalizable}\
 \\
 \hline
 \ea \ \ \ \ \ \ \ \ \ \ \  \  \  \ \ \ \ \ \
  \ \ \ \ \ \ \ \ \ \ \ \ \ \ \ \ \ \ \  \  \  \ \ \ \ \ \
  \ \ \\
   \ea
   \\
   \\
 \label{THS}
 \een
By the blobs we mark here the three main constructive CCM-THS steps.
In the first two steps (viz., $\bullet$ and $\bullet \bullet$) we
may assume to stay inside the usual CCM framework in which the
ground-state eigenvector $|\Psi\kt$ of the quantum system in
question is reparametrized in terms of operator $S$. Thus, the
CCM-THS innovation only emerges, via operators $\tilde{S}$ {\em
alias} $\Theta$, in the third step ($\bullet \bullet \bullet$, see
Table \ref{tab:widebe}).

In this setting let us remind the readers that the (certainly, in
general, existing) creation-operator components of $\Theta^{(CCM)}$
may be expected to play just a marginal role in the convergence. The
reason is that the CCM choice of $\Omega=\exp S$ is mainly aimed at
the construction of the many-body ground states. Thus, a lot of
freedom is left for the introduction of more variational parameters
via $\tilde{S}\neq I$. In contrast, the balanced distribution of
attention of the universal THS formulae between the ground and
excited states lowers, certainly, the latter freedom because the THS
recipe {defines} the metric in terms of $\Omega$ unambiguously.

\subsection{Towards the infinite-dimensional Hilbert spaces}

Once we decide to leave the language of computing and once we move
to the exact description of realistic quantum systems and to the
(say, separable) infinite-dimensional Hilbert spaces, the search for
the CCM-THS unification becomes perceivably more difficult. From the
THS perspective, in particular, the key subtlety lies in the fact
that whenever one decides to treat the two topological vector spaces
$ {\cal H}^{(P)}$ and $ {\cal H}^{(F)}$ (naturally, still without
any account of the definition of the inner products and of the
metrics) as distinct, the map $\Omega=\exp\,S$ will change slightly
its meaning as well as its interpretation.

From the alternative (and also historically older) CCM point of view
it is necessary to recall, first of all, the results of the
important paper \cite{bishop6}. Its author accepted the usual,
above-described CCM linked-cluster parametrization, in its most
general time-dependent form, as deduced from an appropriate action
principle. In turn, this enforces a symplectic structure on the
ensuing CCM phase space of the real-valued ``degrees of freedom''
${\cal S}_{\jmath}, \tilde{{\cal S}_{\jmath}}$ of Eq.~(\ref{bar_A}).

At this point the author of Ref.~\cite{bishop6} has been forced to
discuss the emergence of the characteristic non-Hermiticity of the
average-value functionals $\bar{\Lambda}({\cal S}_{\jmath},
\tilde{{\cal S}_{\jmath}})$ of physical observables as well as of
the action $\bar{\cal A}({\cal S}_{\jmath}, \tilde{{\cal
S}_{\jmath}})$ itself.

In fact, our present idea of possible CCM-THS correspondence also
found another source of inspiration in his approach so let us recall
now his key ideas in more detail. Firstly, he introduced the set of
complex conjugate variables  ${\cal S}_{\jmath}^*, \tilde{{\cal
S}_{\jmath}}^*$ and showed how they could be used to enlarge the CCM
phase space into a genuine complex manifold but of too large a
dimensionality. He further showed how the extra degrees of freedom
could then be eliminated via the Dirac bracket technique. A set of
constraint functions was introduced which thereby select the
physical submanifold ({\it alias} the reduced phase space,  or
constraint surface) corresponding to the original Hilbert space.
Subsequently, the reduced phase space was shown to be a (K\"{a}hler)
complex manifold with a symplectic structure, just as the original
extended one.

Ultimately, the K\"{a}hler manifold may be perceived as defining a
positive, invertible, Hermitian geometry in the reduced phase space.
Arponen \cite{bishop6} further shows that for a compound operator
product $Q=\Lambda_1\Lambda_2$, the CCM star product which generates
the expectation-value functional
$\overline{Q}=\overline{\Lambda_1\Lambda_2}$ in terms of the
individual expectation values $\overline{\Lambda_1}$ and
$\overline{\Lambda_2}$, as given by Eq.~(\ref{bar_A}), can be well
defined in the reduced (i.e., physical) phase space.

This result suggests that besides starting from the THS scheme, one
could also try to develop certain innovative and {consequently
Hermiticity-preserving} hierarchical approximation schemes strictly
within the CCM framework. A judicious use of the on-shell star
products seems capable of establishing another form of the CCM-THS
parallels, and of doing so in and entirely general setting. In
addition, some explicit and concrete {\em constructive}
implementations of the concept of the metric $\Theta$ may be found
{\em directly} in the generic CCM framework. Naturally, a deeper
analysis would require a verification in terms of explicit
constructions. Further development of such a project lies,
naturally, beyond the scope of our present paper.

\subsection{Outlook}

Let us summarize that in the general THS framework one is expected
to perform all of the practical computations of physical predictions
inside the ``friendliest''" Hilbert space ${\cal H}^{(F)}$. What is
a real mathematical promise of a search for the new mutual CCM-THS
correspondences is that even the standard probabilistic
interpretation of many-body wave functions need not require a return
to the ``unfriendly'' space ${\cal H}^{(P)}$. In all respects it
becomes easier to replace the latter space by its (unitarily)
equivalent alternative ${\cal H}^{(S)}$. The reason is that the
latter Hilbert space only differs from its more friendly predecessor
${\cal H}^{(F)}$ by an {\em ad hoc} amended inner product.

Our present brief outline of a few explicit CCM-THS correspondences
centered around the fact that  operator $\tilde{S}$ of the CCM
formalism coincides with the Hilbert-space metric operator $\Theta$
after a ``translation of notation'' to the THS-representation
language of Ref.~\cite{SIGMA}. On the background of this comparison
the main potential innovation of the CCM was found in the THS-based
possibility of distinguishing between the {\em three} separate
Hilbert spaces ${\cal H}^{(P)}$, ${\cal H}^{(F)}$ and ${\cal
H}^{(S)}$ which would represent {\em the same} quantum many-body
system.

The change of perspective revealed several CCM-THS parallels as well
as differences. Among the parallels, one of the most inspiring ones
seems to lie in the emerging structural similarity between the CCM
constructions and their IBM (= interacting boson model)
counterparts. The project of our future development of such a
CCM-IBM correspondence seems promising. In the language of physics
it might enable us to keep the initial physical $P-$superscripted
Hilbert space as fermionic while rendering the other two, $F-$ and
$S-$superscripted Hilbert spaces, strictly in the generalized IBM
spirit, carriers of another, generalized (e.g., pseudo-bosonic)
statistics.

In the opposite direction, also the traditional IBM constructions of
effective Hamiltonians could find some new inspiration in their CCM
analogues. In particular, the prospects of a simplification mediated
by the non-unitary invertible mappings $\Omega =\exp S$ need not
necessarily stay bound by their traditional bosonic-image IBM
restrictions. A new wealth of correspondences may be expected to
become implementable between the auxiliary Hilbert space ${\cal
H}^{(F)}$ and the, by assumption, prohibitively complicated physical
Hilbert space ${\cal H}^{(P)}$ (hence, the superscript $^{(P)}$ may
also mean ``prohibitive'').

Ultimately, the technically most productive idea may be seen in the
exceptional role of the  $F-$superscripted Hilbert space in which
the absence of immediate physical interpretation (say, of the
measurable aspects of coupled clusters) appears more than
compensated by the optimal suitability of this particular
representation space for calculations of the, typically, variational
CCM type.

\subsection*{Acknowledgements}

The participation of MZ was supported by the GA\v{C}R grant Nr.
P203/11/1433.

\end{document}